# HYDRO-THERMAL FLOW IN A ROUGH FRACTURE

A. Neuville, R. Toussaint, J. Schmittbuhl
*Institut de Physique du Globe de Strasbourg, CNRS-UMR7516, 5 rue René Descartes, 67084 Strasbourg Cedex*
e-mail: Jean.Schmittbuhl@eost.u-strasbg.fr

**ABSTRACT**

Heat exchange during laminar flow is studied at the fracture scale on the basis of the Stokes equation. We used a synthetic aperture model (a self-affine model) that has been shown to be a realistic geometrical description of the fracture morphology. We developed a numerical modelling using a finite difference scheme of the hydrodynamic flow and its coupling with an advection/conduction description of the fluid heat. As a first step, temperature within the surrounding rock is supposed to be constant. Influence of the fracture roughness on the heat flux through the wall, is estimated and a thermalization length is shown to emerge. Implications for the Soultz-sous-Forêts geothermal project are discussed.

**INTRODUCTION**

The modelling of the fluid transport in low permeable crustal rocks is of central importance for many applications (Neuman, 2005). Among them is the monitoring of the geothermal circulation in the project of Soulz-sous-Forêts, France (Bachler et al, 2003). The transit time of the water into the main fractures between injection and extraction wells has to be carefully estimated in order to provide precise estimates of the experiment duration. Water flux controls both the heat transfer to the fluid and the cooling of the massif (Murphy, 1979; Glover et al, 1998; Tournier et al, 2000). As shown from previous studies (Benderitter and Elsass, 1995; Tournier et al, 2000) only a few important fractures are controlling the fluid exchange between wells. Accordingly the study of the flow within a single fracture appears of central interest. It is also important for large scale numerical simulations of the convective circulation in the massif (Fontaine et al, 2001; Zhao et al, 2003). Up to now, only very simple models for fractures, like the parallel plates model, have been considered, ignoring the impact of the fracture morphology on the fracture transmitivity (Tournier et al, 2000).

In this study, we are interested in the influence of a realistic geometry of the fracture on its hydro-thermal response. Several studies have addressed the permeability of rough fractures (Brown, 1987; Plouraboué et al, 1995; Hakami and Larsson, 1996; Adler and Thovert, 1999; Méheust and Schmittbuhl, 2000, 2001, 2003). Our goal is to show that realistic fracture morphology can be introduced as a perturbation of the parallel plate model and have a strong influence on the heat transport properties even in the limit of laminar flow.

**MODEL DESCRIPTION**

**Aperture model**

A self-affine surface is a possible geometrical model for roughness topographies of fractures as shown in several previous studies (Brown and Scholz, 1985; Schmittbuhl et al, 1993; Schmittbuhl et al, 1995, Bouchaud, 1997). Self-affinity is a scaling invariance property that reads as: $x \to \lambda x$, $y \to \lambda y$, $z \to \lambda^\zeta z$ where $\lambda$ is a scaling factor and $\zeta$ is the roughness exponent. As can be seen the scaling transformation is supposed to be isotropic within the mean plane $(x,y)$ since the scaling factors are the same along $x$ and $y$ axes. The scaling factor, $(\lambda)$ is not the same along the mean plane $(x,y)$ and the out of plane direction $z$, $(\lambda^\zeta)$. For the latter direction, the roughness exponent $\zeta$ describes how anisotropic is the property. Figure 1 shows an example of a synthetic self-affine aperture field $a(x,y)$. In this case, the aperture defined as the volume between two facing fracture surfaces, is also self-affine (Méheust and Schmittbuhl, 2003).

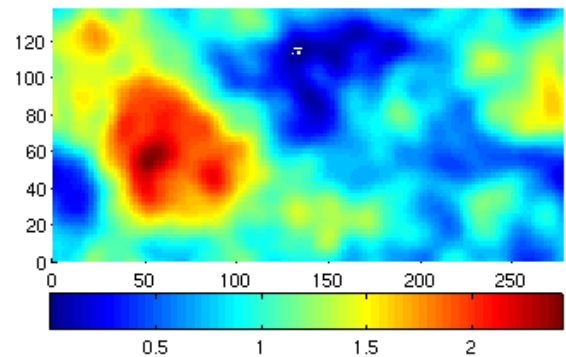

*Figure 1: Sample of a synthetic self-affine aperture field a(x,y) (in mm): red corresponds to largest apertures and blue to quasi-closed areas. The rms of the aperture field is 0.46mm.*

**Fluid flow model**

To study the viscous flow in a rough fracture, we developed a numerical simulation based on the Stokes equation for low Reynolds number (Méheust and Schmittbuhl, 2001; 2003):

$$\vec{\nabla} P = \eta \Delta \vec{v}$$

where $P$ is the fluid pressure, $\eta$ the fluid viscosity and $v$ the Eulerian fluid velocity.

In the limit of small aperture variations ( $\|\vec{\nabla} a\| \ll 1$ ), the lubrication approximation might hold and the local flux per unit length is then related to the local pressure gradient:

$$\vec{u} = -\frac{a^3}{12\eta}\vec{\nabla} P$$

This leads to the Reynolds equation which is a two dimensional approximation of the local flux within the fracture aperture:

$$\vec{\nabla}.(a^3 \vec{\nabla} P) = 0$$

We used a finite difference scheme to solve the Reynolds equation with a conjugated gradient method. Precision was set to $10^{-15}$. Figure 2 shows the result of the computation for the aperture field shown in Fig. 1 when the fluid is injected from the left side. Upper and lower boundary are closed. The fracture is supposed to be horizontal.

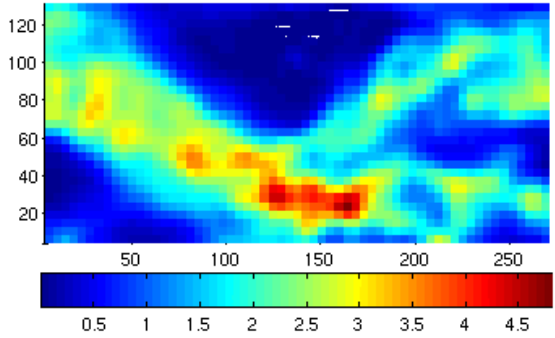

*Figure 2: Map of the normalized local flux for the aperture field shown in Fig. 1. Red areas correspond to maximum fluxes and blue regions to immobile zones.*

It is of important to note the strong channelling of the flow which is induced by the fracture roughness. Despite isotropic aperture fluctuations along the mean fracture plane, as shown in Fig. 1, most of the flow occurs in a significant sub-part of the fracture. This localization of the flow stems from the long range correlations included in the aperture fluctuations. A direct consequence is the clear separation between hydraulically active and inactive zones with a non-direct mapping between aperture magnitude and local flux. In other terms, at a mesoscopic scale, mechanical and hydraulical apertures might strongly differ even if the average aperture is well defined and constant over the whole fracture.

**Heat flow model**

We include heat flow in our modelling to describe the temperature field within the fracture when the fluid is injected at a different temperature from the surrounding rock. $T_0$ is the temperature of the fluid at the inlet of the fracture and $T_w$ is that of the rock. As a first step in our modelling, we assume the temperature of the rock to be homogeneous. We also consider a stationary condition.

Conduction describes the heat flow between two regions of different temperature and fulfills the Fourier law: $\vec{q} = -\lambda \vec{\nabla} T$ that relates linearly the heat flux $q$ and the temperature gradient by the mean of the thermal conductivity $\lambda$. Convection is due to fluid motion and it is described by the estimate of the advected heat. From the energy conservation law (Landau and Lifchitz, 1994), conduction effects balance convective effects and the local temperature within the fluid fulfills:

$$\chi \Delta T = \vec{v}.\vec{\nabla} T$$

where *v* is the fluid velocity and $\chi$ the thermal diffusivity.

For the fluid flow modelling we based our approach on the lubrication approximation which relies on an averaging procedure over the thickness of the aperture. Similarly, we propose a "thermal lubrication" approximation which is based on the average of the temperature over the fracture thickness, weighted by the local fluid velocity:

$$\bar{T}(x,y) = \frac{\int_a v(x,y,z) T(x,y,z) dz}{\int_h v(x,y,z) dz}$$

We also introduce the Nusselt number $Nu = -j_w / q_{ref}$ which compares, when no convection holds, two estimates of the conductive flux: the local flux at the boundary of the fracture $j_w = -\lambda \left(\frac{\partial T}{\partial z}\right)_{z=a/2}$ and the mesoscopic flux at the fracture thickness scale $q_{ref} = \lambda \frac{T_w - \bar{T}}{a}$.

Finally, assuming that the horizontal conduction is negligible when compared to the vertical conduction and that the vertical convection is limited, the local average temperature $\bar{T}(x,y)$ is obtained from the following equation:

$$a\vec{v}.\vec{\nabla} \bar{T} + 2\frac{\chi}{a} Nu(\bar{T} - T_w) = 0$$

We discretized this equation using a first order finite difference scheme and solved it using also a conjugated gradient method. Fig. 3 shows the fluid temperature field for the fracture aperture described in Fig. 1 and in horizontal position. The cold fluid is injected from the left side within the fracture which is assumed to be embedded in hot rock at a homogeneous temperature.

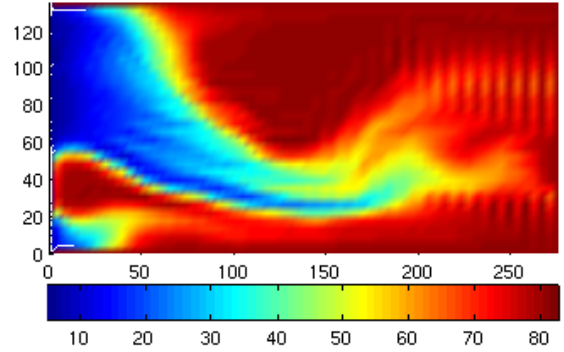

*Figure 3: Map of the normalized temperature within the fracture shown in Fig. 1. The surrounding rock is hotter than the injected fluid. Low temperature zones are in blue and hot zones in red.*

The channelling effect of the temperature field is very significant and directly related to the channelling of the fluid flow (for a convection dominated regime). Accordingly the temperature is far from homogeneous in the fracture though its aperture is constant on average. It therefore emphasizes the importance of aperture fluctuations on the heat exchange between the cold fluid and the hot rock.

Influence of the fracture roughness can be monitored from the comparison with the parallel plate case with the same average aperture, *i.e.* same mechanical aperture. The temperature profile along the fluid flow direction (x direction) can be analytically computed in the case of parallel plate boundary conditions. It reads as:

$$\bar{T} - T_w = (T_0 - T_w)\exp(-x/l_{ref}) \text{ where } l_{ref} = \frac{Pe}{Nu}.$$

The Peclet number is defined as: $Pe = \frac{\bar{v} a}{2\chi}$. Therefore a characteristic length exists in the problem: $l_{ref}$. It describes the typical length scale for the fluid to be in thermal equilibrium with the surrounding rock. Accordingly this is a thermalization length. Figure 4 shows the temperature profile for the parallel plate model and its comparison with profiles for rough fractures. It is of interest to note that rough fractures still show a similar trend but the thermalization length is significantly larger (smaller increase of the temperature within the fracture). In other words, cold fluid migrates further into the fracture in the rough case compared to the parallel plate case owing to the channeling effect.

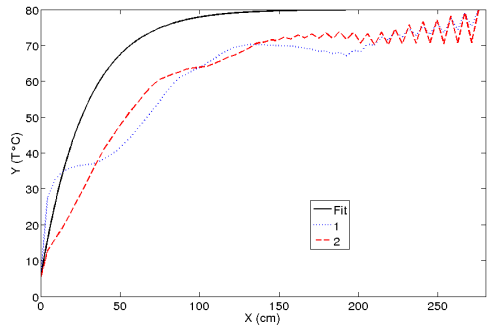

*Figure 4: Temperature profiles (T(x)) for the parallel plate case (exponential curve in black) and two different synthetic fractures (blue and red curves).*

At large scales, numerical oscillations emerge which stems from a too loose spatial discretization of the thermalization length scale.

**CONCLUSIONS**

We propose a numerical model to estimate the heat exchange at the fracture scale between a cold fluid and hot surrounding rock. The numerical model is based on a lubrication approximation for the fluid flow and a "thermalization lubrication" approximation for the temperature evolution. Gravity is neglected since we are dealing with horizontal fractures but could be included from a redefinition of the pressure to include hydrostatic contributions.

Our model shows that fracture roughness is responsible for channelling effects. Fluid flow is dominant in a significant subpart of the fracture where head advection is important. Accordingly, temperature distribution is strongly affected by small fluctuations of the fracture aperture.

On-going work addresses implications for the Soultz-sous-Forêts HDR project where large sub-horizontal fractures are controlling the fluid transfer from the injection well (GPK3) and the extraction wells (GPK2 and GPK4). For this purpose we link our fracture model to a large scale dipolar model of the fluid flow.